\documentclass[aps,prb,twocolumn,showpacs,groupedaddress,superscriptaddress,floatfix]{revtex4}  

\usepackage{graphicx,color}

\newcommand{\be}{\begin{equation}}
\newcommand{\ee}{\end{equation}}
\newcommand{\bea}{\begin{eqnarray}}
\newcommand{\eea}{\end{eqnarray}}

\newcommand{\p}{\partial}
\newcommand{\ri}{{\rm i}}
\newcommand{\re}{{\rm e}}

\newcommand{\EQ}{\begin{equation}}
\newcommand{\EN}{\end{equation}}

\begin{document}
\title{Scaling of excitations in dimerized and frustrated spin-1/2 chains}
\author{D. Controzzi}
\affiliation{International School for Advanced Studies and INFN, Via Beirut, 1, 34100, Trieste, Italy}
\author{C. \surname{Degli Esposti Boschi} }
\affiliation{CNR-INFM, Research Unit of Bologna, Viale Berti-Pichat, 6/2, 40127, Bologna, Italy}
\author{F. Ortolani}
\affiliation{Physics Department, University of Bologna and INFN, Viale Berti-Pichat, 6/2, 40127, Bologna, Italy}
\affiliation{CNR-INFM, Research Unit of Bologna, Viale Berti-Pichat, 6/2, 40127, Bologna, Italy}
\author{S. Pasini}
\affiliation{Physics Department, University of Bologna and INFN, Viale Berti-Pichat, 6/2, 40127, Bologna, Italy}

\begin{abstract} 
We study the finite-size behavior of the low-lying
excitations of spin-1/2 Heisenberg chains with dimerization and 
next-to-nearest neighbors interaction, $J_2$. The numerical analysis, 
performed using  density-matrix renormalization group,  confirms
previous exact diagonalization results and shows that, for different
values of the dimerization parameter $\delta$, the elementary triplet and singlet
excitations present a clear scaling behavior in a wide range of
$\ell=L/\xi$ (where $L$ is the length of the chain and $\xi$ is the
correlation length). At $J_2=J_{\rm 2c}$, where no logarithmic
corrections are present, we
compare the numerical results with finite-size predictions for the sine-Gordon
model obtained using L{\"u}scher's theory. For small $\delta$ we find 
a very good agreement for $\ell
\gtrsim 4$ or $7$ depending on the excitation considered. 
\end{abstract}


\maketitle

Spin-1/2 chains with frustration and explicit dimerization have attracted a
lot of attention for their experimental relevance
to spin-Peierls materials \cite{exp}  
as well as for their interesting theoretical properties, related, for
instance, to the 
presence of two independent mechanisms for spin-gap generation. 
The difficulties in studying these systems are a 
consequence of the lack of quantitative 
analytical methods to describe their full phase
diagram and of different types of limitations in 
the currently
available numerical methods. On the one hand, with exact diagonalization one
can treat at most few dozens of sites, which 
is inadequate when the correlation length becomes large.
On the other, using the density-matrix renormalization group (DMRG) 
technique \cite{Wh93} 
it is possible to reach hundreds of sites, but the control
on the quantum numbers of the excitations is much more 
complicated. 
In both cases it is difficult
to reconstruct the infinite-size spectrum in absence of a theory that
describes the finite-size effects, especially
because some excitations (like the singlet - see below) are known to have
a {\it non-monotonic} dependence on the chain length $L$
(Ref. \onlinecite{japaridze}).
In this paper we present 
a detailed finite-size analysis of DMRG results for 
the low-energy excitations of a spin-1/2 chain with frustration and 
explicit dimerization,  with the aim of getting a deeper theoretical 
understanding of
the corrections due to finite $L$. In particular, along a specific line in 
parameter space, where the spin chain is described by a massive  integrable
quantum field theory (QFT), we compare the numerical results with analytical 
predictions for the finite-size scaling based on  L{\"u}scher's 
theory \cite{luscher,KM}. 

The Hamiltonian of a frustrated spin-1/2 chain with explicit dimerization 
has the standard form
\be
\label{H}
H=J \sum_{i=1}^L \left [ (1+(-)^i \delta) \right ] {\bf S} _i  \cdot
{\bf S}_{i+1} +J_2   \sum_{i=1}^L {\bf S}_i  \cdot
{\bf S}_{i+2}
\ee 
(periodic boundary conditions are assumed in the following).
For $\delta =0$ one identifies two regimes. For 
$J_2<J_{\rm 2c}\cong 0.24 J$ the
model is gapless and described, at low energies, by the $SU(2)$
Wess-Zumino-Witten-Novikov (WZWN) conformal field theory at level 
$k=1$ ($SU(2)_1$), with a 
marginally irrelevant perturbation. 
When $J_2 > J_{\rm 2c}$ the
marginal perturbation changes sign and induces a gap in the spectrum. 
In terms of operators of the WZWN theory the dimerization 
term is proportional to the trace of the $SU(2)$ matrix field $g$. Then,
in the low energy limit, 
(\ref{H}) is described by the following Hamiltonian
density \cite{ah,affleck}
\be
\label{Heff}
{\cal H}={\cal H}_{SU(2)_1} - \gamma~  \bar{\bf J} \cdot {\bf J}
 +\eta ~ {\rm Tr} g,
\ee
where ${\cal H}_{SU(2)_1}=\frac{2\pi v}{3} \left [:{\bf J} \cdot {\bf J}:
  +:\bar{\bf J} \cdot \bar {\bf J}: \right ]$, ${\bf J}$ and $\bar{\bf J}$ are
the chiral $SU(2)$ currents and satisfy the level $k=1$ Kac-Moody
algebra. The
coupling constants are linearly related to $J_2$ and $\delta$: $\gamma \propto
 \left (
1- J_2/J_{\rm 2c} \right )$ and $\eta \propto \delta$.
It should always be kept in mind that, roughly speaking, the field
theory description is valid if the masses are much smaller then the
bandwidth, that implies $\delta \ll 1$ or, for $\delta=0$,
$(J_2-J_{\rm 2c})/J  \ll 1$.

The theory (\ref{Heff}) is integrable when one of the two coupling constants
vanish. For $\eta=0$  excitations are massive (massless)
spinons for $\gamma<0$ ($\gamma>0$ ). When $\eta\neq 0$, the strongly 
relevant perturbation
${\rm Tr} g$ introduces a confining interaction between the spinons that are
no longer fundamental excitations of
(\ref{Heff}) \cite{affleck.confinement,CM}.
For $\gamma=0$ the model is again integrable and equivalent to the
sine-Gordon (SG) model, ${\cal L}=1/2 (\p _\mu \varphi)^2 - g_1/(2 \pi^2 v)
\cos(\beta \varphi)$, at $\beta^2=2\pi$, and the spectrum contains
triplet and singlet excitations\cite{affleck, zam.zam}. 
The ratio between their masses, 
$R=m_{\rm s}/m_{\rm t}$, is known exactly to be $R=\sqrt{3}$. 

The behavior of $R$ for finite values of $\gamma$ (and $\eta\neq 0$) is
difficult to address analytically  even in the 
regime of validity of the field theory description,
because (\ref{Heff}), like its lattice counterpart (\ref{H}), 
is not integrable.  
This problem was studied using exact diagonalization in
Ref. \onlinecite{japaridze},   
where it was found that the singlet is stable for any 
$J_2 < J_{\rm 2c}$ and $R$ interpolates between $\sqrt{3}$ at $J_2=J_{\rm 2c}$
and $R=2$ at $J_2=0$. These results are reproduced by our DMRG
analysis, although we do not report them here (see however
Ref. \onlinecite{CV_etal} 
for the extension to $J_2 < 0$).
Some analytical predictions can be made in the
vicinity of the two integrable points, where a perturbative
analysis can be carried on \cite{affleck.confinement,CM}.

For the analysis of the above problem as well as many others it is essential
to have a theory for the finite-size scaling that allows one to extrapolate
numerical results. For massive integrable QFTs it is possible to express the
leading finite-size corrections to the spectrum in terms of their exact
scattering data\cite{luscher,KM}. As we recalled, there are two points
for which (\ref{Heff}) is integrable. 
In this paper we consider the integrable point $\gamma = 0$ and $\eta \neq 0$,
rather than $\eta = 0$ and $\gamma < 0$, because the former case presents bound states
and is physically more interesting. Numerically, this is also the easiest scenario to investigate 
because in the other case the gap opens very weakly \cite{CPKSR}.
Hence, in the following we will compare DMRG
data for the lattice Hamiltonian (\ref{H}) with $J_2=J_{\rm 2c}$ and
$\delta \ll 1$ with predictions coming from the SG theory at $\beta^2 = 2 \pi$.

Let us start presenting the numerical results. Before studying the
finite size-effects we extract from the numerical data some important
non-universal pre-factors -- a problem recently addressed in
Ref.~\onlinecite{Or}. It is well known that, in absence of logarithmic
corrections, the triplet mass scales as $m_{\rm t} = A_{\rm t}
\delta^{2/3}$, with $A_{\rm t}$ being a non-universal amplitude.
The energy gaps, computed with DMRG,  are linearly related to the masses
through a velocity pre-factor that we express as $v = C_v J \pi/2$:
$\Delta E_{\rm t}=v m_{\rm t}$ (the lattice spacing is formally set to one).
For $J_2 = 0$ it is known that $C_v = 1$, but
for $J_2 = J_{\rm 2c}$ the constant may assume a different value.
Another unknown constant connects $\delta$ with the coefficient of the 
relevant term in the SG model: 
$g_1 = 6 J \delta (\pi/2)^{1/4} C_\delta$ 
($C_\delta=1$ for $J_2=0$ \cite{Or}).
Along the same lines of Ref.~\onlinecite{Or} we can extract these constants from the
pre-factors of the 
triplet gap and ground-state energy density, that we write as $\Delta E_{\rm t}/J = 1.723 K_{\rm t}
\delta^{2/3}$ and $[e_{\rm GS}(\delta)-
e_{\rm GS}(0)]/J= -0.2728 K_{\rm GS} \delta^{4/3}$, respectively.
The relation between the constants is $K_{\rm t} = C_v \left(C_\delta/C_v\right)^{2/3}$ and 
$K_{\rm GS} = C_v \left(C_\delta/C_v\right)^{4/3}$.
We have computed these quantities for some values of $\delta$ using
the finite-system DMRG algorithm \cite{Wh93} with three ``zips'' in
a range of $L$ from 10 to 100. The results are presented in Table
\ref{tab1}, and imply $C_v \simeq 0.81$ and $C_\delta \simeq 1.53$
at $J_2=J_{\rm 2c}$,
to be compared with $C_\delta=C_v=1$ at $J_2=0$.

\begin{figure}
\begin{center}
\includegraphics[height=50mm,keepaspectratio,clip]{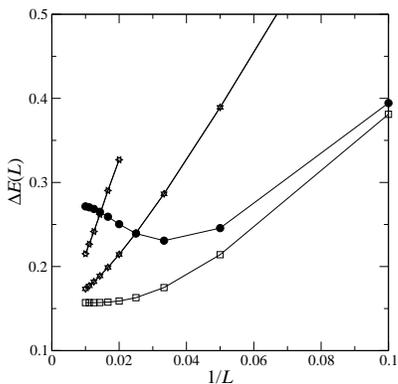}
\caption{Finite-size gaps vs $1/L$ for $\delta=0.02$ and $J_2 = J_{\rm 2c}$: 
triplet (open squares), singlet (full circles), lowest 1P triplet excitations 
with momenta different from $0$ and $\pi$ (coinciding up and down triangles). 
Following the singlet at large $L$ requires to target various excited states 
within $S^z_{\rm tot}=0$ ($N_0=6$ in this example, where $M=256$), due
to the crossing with 1P states. The singlet curve has a minimum at $L_{\rm min}(0.02) \simeq 30$.}
\label{del002alc}
\end{center}
\vspace{-6mm}
\end{figure}

Let us now come to the central part of the paper by examining the finite-size
features of the low-energy spectrum. In the framework of the DMRG this
is done by building the superblock density matrix as a mixture
(with equal weights) of the matrices associated with a prescribed
number of target states. By means of the thick-restart variant
of the Lanczos method \cite{MtDMRG} we are able to target up to ten excited
states 
in each sector of $S_{\rm tot}^z = \sum_i S_i^z$, which is the only
good quantum number that we have exploited.
First of all from the so-called
1-particle (1P) excitations of the triplet we can get an independent estimate
of the velocity. In fact, these excitations for small momenta 
$q_n = 2 \pi n/L$ (with $n=\pm 1, \pm 2,\dots$) have energy $E_n = E_{\rm GS} + \sqrt{\Delta E_{\rm t}^2+v_1^2 q_n^2}$.
So we have checked that $L \sqrt{\Delta E_n^2-\Delta E_{\rm t}^2}/2 \pi$ actually scale
as $n v_1$ and we have reported the values of $v_1$ in Table
\ref{tab1}.
Unfortunately, these states with nonzero momentum represent also the main limitation
that we have encountered in the computation of the singlet gap $\Delta E_{\rm s}$ at 
large values of $\delta$. This is because all
the excited states that we can target in $S^z_{\rm tot}=0$, say $N_0$, are eventually exhausted by
the triplet and its 1P excitations. In practice, the singlet 
``disappears'' from the set of DMRG levels for $L \gtrsim (N_0-2) \pi v_1/\Delta E_{\rm t} \sqrt{R^2-1}$,
where $R$ is the mass ratio \cite{Lmax}. 
In addition, unlike the triplet, the singlet level
has a non-monotonic dependence on $L$. Typically $\Delta E_{\rm s}(L)$ decreases
with increasing $L$ up to a characteristic value $L_{\rm min}(\delta)$, beyond which
it starts to increase and converges to the asymptotic limit {\it from below}.
An example of such a feature is plotted in Fig. \ref{del002alc} for
$\delta = 0.02$ and $J_2 = J_{\rm 2c}$.
As listed in Table \ref{tab1} the values of $L_{\rm min}$ increase for $\delta \to 0$
so that there exists a minimum value of $\delta$ below which $L_{\rm min}$ becomes
larger than the maximum reliable size that we can handle with the DMRG.

\begin{table}
\centering{
\begin{tabular}{|l|c|l|l|c|c|}
\hline
$\delta$ & $\Delta E_{\rm t}/J$ & $K_{\rm t}$ & $K_{\rm GS}$ & $v_1/J$ & $L_{\rm min}(\delta)$ \\
\hline
\hline
0.01 & 0.09875 & 1.235 & 1.966 & 1.1921 & 50 \\
0.02 & 0.15705 & 1.237 & 1.948 & 1.1822 $\pm$ 0.0010 & 30 \\
0.10 & 0.46233 & 1.245 & 1.800 & 1.1425 $\pm$ 0.0003 & 10 \\
\hline
\end{tabular}
\caption{Triplet gaps, constants $K_{\rm t}$, 
$K_{\rm GS}$ and estimates of the 
``1-particle velocity'' $v_1$  for different values of $\delta$
and $J_2 = J_{\rm 2c}$ 
(see text after Eq. (\ref{devtl}) for the determination of $\Delta
E_{\rm t}/J$.).  
Typically $M=256$ optimized
DMRG states have been retained. When $\Delta E_{\rm t} < 0.1$ 
we have improved the precision using $M=405$.
The last column contains the (approximate) values at which the 
singlet gaps attain a minimum.}
\label{tab1}
}
\vspace{-6mm}
\end{table}

There are other cases in the literature for which numerical results 
show that some states in the
spectrum exhibit a non-monotonic dependence on $L$ (see for instance
Refs. \onlinecite{ZWXSY}).
In such situations it is particularly difficult  to get a reliable estimate
of the infinite-size value. In some cases it is possible to push $L$
beyond the minimum and even beyond the
flex where the finite-size data start to saturate. Whenever this is
not the case the construction of a 
suitable fitting function becomes essential. 
In Ref.s \onlinecite{japaridze,BRDS}, some empirical 
scaling functions were suggested 
that provide a good fit of the numerical data. Nevertheless these expressions
were not obtained from a consistent theoretical construction and contain various
fitting parameters. 
In the following we will construct the finite-size corrections using the QFT
description. In particular we will compute rigorously 
the dominant contribution to the scaling functions for large $L$. These
expressions contain no fitting parameters.

As already recalled for $J_2=J_{\rm 2c}$ and $\delta \ll1$ the spin chain
is described by the integrable 
SG model at $\beta^2=2\pi$. 
The factorized scattering theory explicitly shows the hidden
$SU(2)$ symmetry of the model at  this point \cite{affleck}. 
The spectrum consists of a soliton $s$ and an anti--soliton $\overline s$ 
which are degenerate with a breather state $b_1$. Moreover, 
all these particles have the same S-matrix \cite{zam.zam,affleck}
\be
\label{Smat1}
S_{a,b}(\theta)=S_0(\theta)=
\frac{\sinh \theta +{\rm i}\sin \pi/3}{\sinh \theta -{\rm i}\sin \pi/3}
, ~~a,b=s,\bar s, b_1 \;.
\ee
The rapidity $\theta$ parameterizes the relativistic dispersion
relations: $e_a=m_a \cosh \theta$, $p_a=m_a \sinh \theta$.
In addition, there is another 
breather state $b_2$ of higher mass.
The additional S-matrices are \cite{zam.zam,affleck}
\bea
&&S_{a,b_2}(\theta)=S_0(\theta+{\rm i}\pi/6) S_0(\theta-{\rm i}\pi/6)
\, ,
~~a=s,\bar s, b_1\, ; \;\;\;\;
\label{Smat2}
\\
&&S_{b_2,b_2}(\theta)=\left [ S_0(\theta) \right ] ^3.
\label{Smat3}
\eea
The excitations 
can be then organized into a triplet ($s,\bar s, b_1$) of 
mass $m_{\rm t}$ and a singlet of mass $m_{\rm s} = \sqrt{3} m_{\rm t}$.

\begin{figure}
\begin{center}
\includegraphics[height=20mm,keepaspectratio,clip]{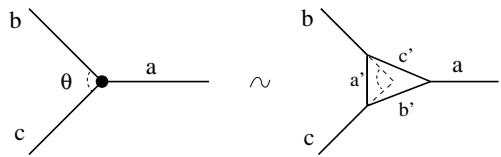}
\caption{Physical interpretation of third order poles. The angle $\theta$
  is related to the poles in the intermediate processes as
  $\theta=u_{b,a'}^{c'}+u_{c,a'}^{b'}-\pi$.}
\label{fig:poles}
\end{center}
\vspace{-6mm}
\end{figure}

For what follows, it is worth recalling that bound states, like the
breathers, are
associated to poles in the S-matrix. In particular if 
$\theta=\ri u_{bc}^a$ is a pole in the scattering process of the
particle $b$ and $c$ the mass of the bound state is given by:
$m_a^2=m_b^2+m_c^2+2 m_bm_c \cos u_{bc}^a$. For instance, 
the simple poles in the
S-matrix $S_{s,\bar s}(\theta)$ at $\theta=2 \ri \pi/3$ and $\theta=\ri \pi/3$
correspond to the two breathers (it is easy to check their
masses are $m_{\rm t}$ and $m_{\rm s}$ respectively). The third order
pole at $\theta=2\pi i/3$ in $S_{b_2,b_2}(\theta)$ should be interpreted as
associated to intermediate virtual particles in the scattering
processes \cite{triple.poles}, as drawn in Fig.~\ref{fig:poles}. 
All other poles in (\ref{Smat1}-\ref{Smat3}) are redundant and do not
correspond to additional bound states \cite{zam.zam}.

The finite-size corrections for a QFT in a large but finite volume with
periodic boundary conditions are  a consequence of the vacuum polarization
and, when the model is integrable, they  can be extracted
from the exact scattering data of the infinite-volume theory \cite{luscher,KM}.
The leading corrections to the 
masses are exponentially small and consist of two terms
\be
\label{fplusmu}
\Delta m_a(L) =
\Delta m_a^{(F)}(L) +\Delta m_a^{(\mu)}(L),
\ee
which roughly can be understood as follows. The first term exists in
any theory and can be interpreted as virtual particles 
``traveling around the world'' once before being annihilated
again. It has the form 
\be
\label{Ft}
\Delta m_a^{(F)}(L)=-\sum_b \int_{-\infty}^{\infty} \frac{{\rm d}
  \theta}{2\pi} {\rm e}^{-e_b(\theta) L }  
e_b(\theta) f_{a,b}(\theta) \; ,
\ee
where $
f_{a,b}(\theta)=  S_{a,b}(\theta+\frac{\ri \pi}{2})-1$. 
For large $L$ Eq.~(\ref{Ft}) is of order ${\cal  O} (\re^{-m_b L})$.
The second term in (\ref{fplusmu}) is present only in theories with bound
states and is associated to virtual processes in which a particle is
split into its two constituents that ``travel around the
world'' before recombining again into the original particle. 
If the particle $a$ is
bound state of two other particles of the theory, say $b,c$, associated
to the pole $\theta=\ri u_{bc}^a$, then $\Delta m_a^{(\mu)}(L)$ reads
\be
\label{mut}
\Delta m_a^{(\mu)}(L) = 
-\sum_{b,c} \Theta \left ( m_a^2-|m_b^2-m_c^2| \right)
\mu_{abc} R_{abc} {\rm e}^{-\mu_{abc} L} \; ,
\ee
where $\Theta$ denotes the Heaviside step function and 
\bea
\mu_{abc}=m_b \sin u_{ab}^c \;,
R_{abc}=-{\rm i} \, {\rm Res}_{\theta={\rm i} u_{ab}^c}
  S_{a,b}(\theta) \;.
\eea
As mentioned before the third order pole has to be interpreted as formation
of virtual particles and $R_{abc}=R_{ab'c'}R_{ba'c'}R_{cb'a'}$. 
In general, since for small $L$, 
$\Delta E_{\rm s,t} (L)\sim 1/L$, 
if the terms above approach the infinite-volume
limit from below the scaling function will present a
minimum (the minimum may also appear as a
consequence of the competition of the two terms in (\ref{fplusmu}) \cite{KM}).

\begin{figure}
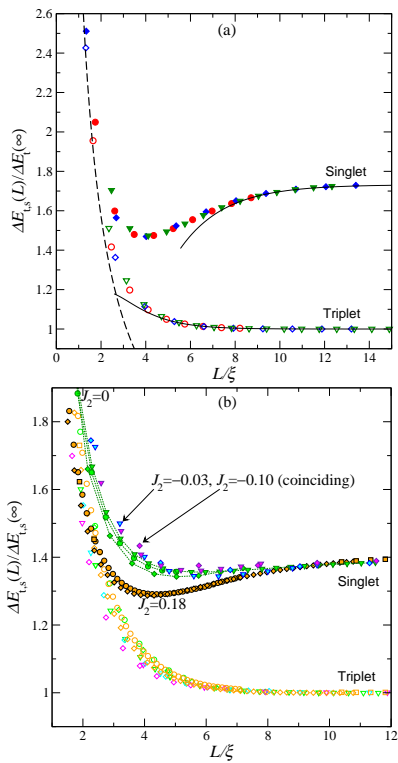

\begin{center}
\includegraphics[height=50mm,keepaspectratio,clip]{sp_al0241.eps}
\includegraphics[height=50mm,keepaspectratio,clip]{sp_val.eps}
\caption{(a) Scaling plot of finite-size corrections to the triplet 
and singlet gaps $\Delta E_{\rm t,s}(L)$ at $J_2 = J_{\rm 2c}$. 
Continuous lines
represent the analytical predictions  (\ref{devtl}) and
(\ref{devsl}), while
the symbols refer to DMRG data obtained with the
same parameters as in Table \ref{tab1} (circles: $\delta = 0.01$; diamonds:
$\delta = 0.02$; triangles: $\delta = 0.10$). The 
dashed line shows the $1 /\ell$ behavior at small $\ell$. 
(b) Same as in (a) for $J_2 < J_{\rm 2c}$ (squares: $\delta = 0.06$).
The singlet data have been rescaled by $\Delta E_{\rm s}(\infty)$ and then shifted by +0.4 for clarity reasons.
}
\label{sp}
\end{center}
\vspace{-6mm}
\end{figure}

Since the singlet and the triplet masses are related, 
all the finite-size corrections 
depend on the scaling variable $m_{\rm t} L$
(Ref. \onlinecite{comment.regularization}). Using formulas
(\ref{fplusmu}-\ref{mut}) and the pole
structure of the S-matrix, we find the following expression for the 
leading finite-size corrections,
$ \Phi_{\rm s,t}(\ell)\equiv \Delta E_{\rm s,t}(L) / \Delta E_t(\infty) $,
\bea
&&\Phi_{\rm t}(\ell) = 1 + 6 {\rm e}^{-\ell \sqrt{3}/2} 
-3 \int_{-\infty}^{\infty}\frac{{\rm d}\theta}{2 \pi} {\rm e}^{-\ell
  \cosh \theta} f_0(\theta) \, ,\;\;\;\;
\label{devtl}
\\
&&\Phi_{\rm s}(\ell) = \sqrt{3} \left(
1 - 3 {\rm e}^{-\ell/2} -36 \re^{3\ell /2} \right) \; , 
\label{devsl}
\eea
with $f_0(\theta) \equiv 2 \sqrt{3} \cosh \theta/(2 \cosh \theta -\sqrt{3})$. 
These scaling functions are compared to the scaled DMRG data in 
Fig.~\ref{sp}a
(one has to replace $m_{\rm t} L$ 
with the scaling variable $\ell = L/\xi$). 
For each value of $\delta$ 
the scale factors $\xi$
(slightly different for triplet and the singlet)
and  $\Delta E_{\rm t}$ 
have been tuned in order to give the best collapse of the curves.
Using the numbers found for the triplet, the product 
$v = \xi \Delta E_{\rm t}$
yields another estimate of the velocity, $v = 1.19 \pm 0.01$, in agreement with those 
in Table~\ref{tab1}. 

A first interesting result of this analysis is that stable excitations 
of the lattice model
for $J_2 = J_{\rm 2c}$ with a dimerization term exhibit
a clear scaling behavior over 
the whole range of $L$ for different values of $\delta$.
Moreover, the scaling functions obtained using the QFT 
description
capture well the finite-size effects of the spin Hamiltonian for $\ell
\gtrsim 4$ or  $7$,  
depending on which of the two excitations we consider.
We recall that the analytic results are the exact expression of the leading
finite-size corrections and contain no additional fitting parameters.

Finally, we have repeated the numerical analysis for other
values of $J_2$. Unfortunately in this case
the low-energy effective field theory is no longer integrable and 
computations of finite-size corrections are much more difficult. 
In principle both the
variation of the mass and of the S-matrix can be taken into account
perturbatively \cite{dms}. The numerical results are reported in
Fig. \ref{sp}b. We note that the form of the scaling function for the singlet 
has a much stronger dependence on $J_2$ than the triplet one. For fixed $J_2$ we find again no 
appreciable dependence on~$\delta$. 

We would like to thank G. Mussardo for suggesting to compare the DMRG
results with L{\"u}scher's theory and for essential
discussions. D.C. thanks A. Nersesyan for important discussions and
interest in the work. 
We also acknowledge useful observations made by  E. Ercolessi, G. Morandi,
M. Roncaglia, L. Campos Venuti and  G. Delfino.
This work was partially supported by
the TMR network EUCLID (HPRN-CT-2002-00325), and by the Italian MIUR
through COFIN projects (prot. n. 2002024522\_001 and 2003029498\_013).


\begin{thebibliography}{99}
\bibitem{Lmax} This formula stems from equating $\Delta E_{\rm s} = R \Delta E_{\rm t}$
to the energy of the highest 1P state with momentum $|q|=2 \pi/L (N_0-2)/2$, given
that within the $N_0$ states we have to include the singlet, the triplet and
all the 1P excitations that are doubly degenerate in momentum.
\bibitem{comment.regularization}  
It should be noticed that the
integrals in (\ref{Ft}) involving $f_{a,b_2}(\theta)$, 
with $a$ belonging to the triplet,
diverge and need to be regularized. Such divergences are possible in one
dimensional models, but there is no general prescription to treat 
them \cite{KM}.
We regularized them by simply subtracting the divergent part. In
any case, it turns out that the contribution from this term (not shown
below) is very
small and then the choice of the regularization scheme does not affect
the analysis.
\end{thebibliography}
\end{document}